\documentclass{vldb}

\usepackage{booktabs} 
\usepackage{amsmath}
\usepackage{amssymb}
\usepackage{comment}
\usepackage{tikz}

\usepackage{verbatim}
\usepackage[T1]{fontenc}
\usepackage{caption}
\usepackage{subcaption}
\usepackage{graphicx}
\usepackage{balance}  
\usepackage{url}
\usepackage{setspace}

\newcommand{\squishlisttwo}{
   \begin{list}{$\bullet$}
       { \setlength{\itemsep}{0pt}      \setlength{\parsep}{3pt}
       \setlength{\topsep}{3pt}       \setlength{\partopsep}{0pt}
      \setlength{\leftmargin}{1.5em} \setlength{\labelwidth}{1em}
       \setlength{\labelsep}{0.5em} } }

\newcommand{\squishlistthree}{
   \begin{list}{$\bullet$}
   { \setlength{\itemsep}{0pt}    \setlength{\parsep}{0pt}
   \setlength{\topsep}{0pt}     \setlength{\partopsep}{0pt}
   \setlength{\leftmargin}{2em} \setlength{\labelwidth}{1.5em}
   \setlength{\labelsep}{0.5em} } }
\newcommand{\squishend}{
\end{list}  }

\newcounter{Lcount}
\newcommand{\squishlist}{
\begin{list}{\arabic{Lcount}. }
{ \usecounter{Lcount}
\setlength{\itemsep}{0pt}
\setlength{\parsep}{0pt}
\setlength{\topsep}{0pt}
\setlength{\partopsep}{0pt}
\setlength{\leftmargin}{2em}
\setlength{\labelwidth}{1.5em}
\setlength{\labelsep}{0.5em} } }

\vldbTitle{Integrating Information About Entities Progressively}
\vldbAuthors{Ben McCamish, Christopher Buss, Arash Termehchy, David Maier}
\vldbDOI{https://doi.org/10.14778/xxxxxxx.xxxxxxx}
\vldbVolume{12}
\vldbNumber{xxx}
\vldbYear{2019}

\begin{document}
\title{Integrating Information About Entities Progressively}

\numberofauthors{4} 

\author{
%
%
\alignauthor Ben McCamish\\
  \affaddr{Oregon State University}\\
  \affaddr{mccamisb@oregonstate.edu}
\alignauthor Christopher Buss\\
  \affaddr{Oregon State University}\\
  \affaddr{bussch@oregonstate.edu}
\alignauthor Arash Termehchy\\
  \affaddr{Oregon State University}\\
  \affaddr{termehca@oregonstate.edu}
\and
\alignauthor David Maier\\
  \affaddr{Portland State University}\\
  \affaddr{maier@pdx.edu}
}

\maketitle


\begin{abstract}
Users often have to integrate information about entities from multiple data sources.
This task is challenging as each data source may represent information about the same entity in a distinct form, e.g., each data source may use a different name for the same person. Currently, data from different representations are translated into a unified one via lengthy and costly expert attention and tuning. Such methods cannot scale to the rapidly increasing number and variety of available data sources. We demonstrate ProgMap, a entity-matching framework in which data sources learn to collaborate and integrate information about entities on-demand and with minimal expert intervention. The data sources leverage user feedback to improve the accuracy of their collaboration and results.
ProgMap also has techniques to reduce the amount of required user feedback to achieve effective matchings.

\end{abstract}


\section{Introduction}
\label{sec:intro}
The information about a given entity is often spread across multiple data sources, therefore, users have to integrate information from several data sources. This task is challenging as each data source may represent information in a distinct form, e.g., each data source may refer to the same entity under a distinct name. Users have to translate their queries to forms that are understandable by underlying data sources. This process is traditionally done by writing a set of rules that takes the query or data organized in one form and translates it to the query or data under another representation. This approach, however, takes a very long time, a great deal of manual labor, and constant expert attention to develop and maintain these rules~\cite{yan2013actively}. One might use supervised learning techniques to match entities, but off-line training data is hard to find for this task. 

Due to the enormous upfront cost of matching and integrating entities, the database community has recognized the need to build on-demand integration systems \cite{yan2013actively}. These systems create an initial set of matching and mapping rules between data sources that may have some inaccuracies, then improve their accuracy using the end users' feedback on the results of their queries. Although these systems reduce the upfront cost of building an integration system, they still have to extract or discover the overall content and structure of all data sources and construct reasonably accurate rules between data sources initially. Nevertheless, the (full) information about the overall structure and content of a data source may \textit{not} be available upfront and it will take some time to obtain them. Also, some database owners may prefer \textit{not} to release such information about their databases, due to privacy concerns. They may instead prefer to deliver information in the traditional method on a query-by-query basis. Furthermore, due to the unprecedented rapid growth in the number of online data sources and databases, creating an initial and reasonably accurate mapping may still take a long time and extensive manual labor. For example, one organization may have access to hundreds or thousands of databases in various formats \cite{DBLP:conf/cidr/DengFAWSEIMO017,DBLP:conf/cidr/Madden17}. Moreover, many data sources and pieces of information inside them might \textit{not} interest users and need \textit{not} to appear in the (initial) rules.

We demonstrate \textit{ProgMap}, an on-demand and progressive system, which enables data sources in a domain of interest to learn to collaborate and integrate information about entities and answer users' queries without any manually built initial matching and mapping between data sources. Users submit their queries to a mediator or a local data source. The local data source will craft and send queries to the other, i.e., external, data sources to retrieve information relevant to the users' queries. There is no initial mapping to help the local data source with accurately translating queries. Nevertheless, external data sources may support using some query languages which one can formulate a potentially inaccurate query without complete information about the schema or content of their data and get some partial answers. Examples of such languages are keyword queries and 
so-called full-text queries, e.g., MySQL {\it match} operator.
Thus, the local data source will submit the users' information need in the form of these queries to external data sources. Each external data source may return a list of entities in response. Because these queries are inherently vague, external data sources may \textit{not} precisely understand the need of the local one and return some non-relevant information or \textit{not} deliver all their relevant data. The local data source integrates the returned information with its own local results and presents it to the user. According to the end user's feedback on the returned result, the local data source will revise its method of formulating keyword queries to find relevant tuples from each external data source. If the external data source accepts user feedback, the local data source also shares the user feedback with them and they may also modify how to return relevant entities.

The local and external data sources revise their strategies of formulating queries and returning entities in a way that creates a balance between \textit{ exploiting} the information gained from the successes of the preceding interactions and \textit{exploring} new ways of crafting and answering queries. 
Over the course of several interactions, the local and external data sources will learn to communicate effectively. Since it may require too much user feedback to construct effective mappings, ProgMap uses some techniques to reduce the amount of feedback. Our empirical studies indicate that ProgMap is able to find and match relevant entities with reasonable effectiveness
after a rather small number of interactions.

\section{Framework}
\label{sec:framework}
For the sake of simplicity, we assume that the information about entities in each data source is stored in a single relational table where each tuple represents information about a different entity.
A \textit{local data source} receives users' queries and communicates with and integrates information about the entities relevant to the query from multiple \textit{external data sources}.  
We use Tables~\ref{example:framework:table:instanceA} and~\ref{example:framework:table:instanceB}, which illustrate fragments of product databases in different companies, as our running example. The local database contains information about various products in relation \textit{Products}. The users of the local database wish to see who sells the given products. This information is stored in an external data source containing the relation \textit{Sellers}. Since databases store the information about the same product in different forms, the local data source has to learn how to properly query the external database to find the companies that sell the respective products, combine or join the results on both databases, and present the final answers to the user.

\textbf{Local Query:}
Each round of communication starts when the user of the local data source submits a query. The local data source may find a set of entities that satisfy this query in its own database. Since there is \textit{not} a global schema that covers information in all data sources, these queries are only over the schema of the local database. We also allow users to submit keyword queries over the local database.

\textbf{External Query:}
After extracting the result of the local query in its dataset, the local data source formulates and submits {\it external queries} to the external data sources in order to extract entities relevant to the local query. 
The local data source, however, does \textit{not} precisely know the mapping between attributes in the local and the ones in the external data sources, therefore, it {\it cannot} submit a fully accurate SQL query to retrieve the relevant entities.
Many data sources, however, accept potentially inaccurate queries, which may return some relevant entities with other non-relevant ones.
For instance, many (online) data sources support keyword queries \cite{yan2013actively}. 
Also, commercial RDBMSs often support the so-called full-text queries that do {\it not} require the complete information about mapping between schemas, e.g., MySQL {\it match} or Oracle {\it contains} operators. To answer these queries, data sources return a ranked list of entities where the highly ranked ones are deemed more relevant. Due to the limited space, we explain only the methods of constructing keyword queries.
Since each entity in the local database may be relevant to a different set of entities in each external data source, the local data source may construct an external query per matching local entity. For instance, given that tuple product \textit{Soda} is in the local answers to a user query over Table~\ref{example:framework:table:instanceA}, the local data source may submit external (keyword) queries \textit{Soda Drinks} or \textit{Drinks}.

\vspace{-1em}
\begin{table}[h!]
    \begin{subtable} {0.5\linewidth}
    \caption{Products}
        \begin{tabular}{l l l}
            \hline
            \hline
            ID & Name & Category\\
            \hline
            $s_1$ & Soda & Drinks\\
            $s_2$ & Beef & Meat\\
            \hline
        \end{tabular}
    \label{example:framework:table:instanceA}
    \end{subtable}
    \begin{subtable}{0.48\linewidth}
        \centering
        \caption{External Queries}
        \begin{tabular}{l l}
            \hline
            \hline
            Query\# & \multicolumn{1}{c}{Query} \\
            \hline
            $g_1$ & `Soda Drinks'\\
            $g_2$ & `Beef Meat'\\
            $g_3$ & `Drinks'\\
            $g_4$ & `Meat'\\
            \hline
        \end{tabular}
        \label{example:framework:table:signalsA}
    \end{subtable}

    \begin{subtable} {.5\linewidth}
    \caption{Sellers}
        \begin{tabular}{l l l l l}
            \hline
            \hline
            $r_1$ & P\_Name & P\_Category & P\_Seller & P\_Price\\
            \hline
            $r_1$ & Pop & Drinks & Kroger & 1\\
            Hamburger & Sandwich & 7/11 & 4 \\
            \hline
        \end{tabular}
    \label{example:framework:table:instanceB}
    \end{subtable}
    \label{example:framework:table:instances}
    \vspace{-0.5em}
    \caption{Products and Sellers database and some external queries}
\end{table}
\vspace{-2em}
\begin{table}[h!]
    \begin{subtable}{.5\linewidth}
        \caption{Querying strategy}
        \centering
        \begin{tabular}{c|c|c|c|c|}
            \cline{2-5}
             & \multicolumn{1}{c|}{$g_1$} & $g_2$ & $g_3$ & $g_4$\\
            \hline
            \multicolumn{1}{|c|}{$s_1$} & 0.4 & 0.1 & 0.5 & 0\\
            \hline
            \multicolumn{1}{|c|}{$s_2$} & 0 & 0.4 & 0.1 & 0.4\\
            \hline
        \end{tabular}
        \label{example:framework:table:strategies:sender}
    \end{subtable}
    \begin{subtable}{.48\linewidth}
        \caption{Answering strategy}
        \centering
        \begin{tabular}{c|c|c|}
            \cline{2-3}
             & \multicolumn{1}{c|}{$r_1$} & $r_2$\\
            \hline
            \multicolumn{1}{|c|}{$g_1$} & 0.8 & 0.2\\
            \hline
            \multicolumn{1}{|c|}{$g_2$} & 0.5 & 0.5\\
            \hline
            \multicolumn{1}{|c|}{$g_3$} & 0 & 1\\
            \hline
            \multicolumn{1}{|c|}{$g_4$} & 0.7 & 0.3\\
            \hline
        \end{tabular}
        \label{example:framework:table:strategies:receiver}
    \end{subtable}
    \label{example:framework:table:strategies}
    \vspace{-0.5em}
    \caption{Local and external strategies}
\end{table}
\vspace{-1.5em}
\textbf{Local Strategy:}
The \textit{local strategy} reflects how the local data source formulates the external queries.
Roughly speaking, it crafts an external query per pair of its local query and one of its matched local entities.
We call each of these pairs an \textit{intent}. The local strategy stochastically maps each intent to the set of possible external queries. 
Using our running example, let $s_1$ and $s_2$ denote the tuples with IDs 1 and 2 respectively, in the local database shown in Table~\ref{example:framework:table:instanceA}. The local data source uses the four external queries in Table~\ref{example:framework:table:signalsA} to find the information related to these tuples in the external data source. Table~\ref{example:framework:table:strategies:sender} shows a sample local strategy used by the local data source. If the local data source wishes to find information related to $s_1$, it will send the external query~$g_1$ with~40\% probability. Since the strategy is stochastic, in each interaction, the local data source may send a different query for the same intent.
Queries that have been more successful in the past are assigned higher probabilities in the strategy. 
This allows the local data source to both \textit{exploit} the external queries that have relatively expressed the intent in the past successfully and \textit{explore} other external queries that have \textit{not} been tried sufficiently frequently and acquire more knowledge for the long-run. 
The local data source maintains a separate strategy per external data source.
We use n-gram features, i.e., contiguous sequences of n terms, of the local query and its locally matched entities of intents and n-gram terms of external queries to materialize the local strategy. 
We also add the schema information, e.g., attribute names, to each n-gram feature of the local entities.
Given an intent, the local data source samples from this strategy sufficiently many times to construct a keyword query with a prespecified number of terms (keywords). The terms in the external queries are initially selected from n-gram features in the local query and its matched entities. 
Initially, the n-gram features of the local entity and query of an intent have higher probabilities of being included in its external queries.
Over time, these probabilities will be updated and more queries will be added to the set of possible external queries as explained in Section~\ref{sec:learning}.

\textbf{External Strategy:}
Each external data source answers the external queries using its \textit{ external strategy}.
It is a stochastic mapping from keyword queries to entities in the external data source.
Our framework also supports external data sources that use a deterministic strategy, e.g., fixed ranking formula.
Table~\ref{example:framework:table:strategies:receiver} illustrates an external strategy for the data source with the data in Table~\ref{example:framework:table:instanceB}, where $r_1$ and $r_2$ are the first and second tuples in the database, respectively. 
In this example, if the external data source gets query~$g_2$, it will return tuples $r_1$ or $r_2$ with equal probability.
The stochastic strategy enables the data source to both exploit and explore.
By sampling the candidate answers according to their probabilities in the external strategy $k$ times, the external data source returns a ranked list of $k$ tuples to the local one. 
When a new external query is received, a new entry is added to the strategy.
We maintain the external strategy using n-gram features of queries and the content and schema of entities in the database.

\textbf{Feedback:}
After retrieving related entities from the external data source for each entity in the local results, the local data source combines the local and external results and presents them to the user. For each entity in the local data source, the local data source creates an entry in the results with information about both the local entity and its related external ones. The user informs the local data source whether some matched entities are relevant. The local data source collects the feedback and computes some effectiveness metric to measure the quality of retrieved information for each entity from an external database. We use the well-known effectiveness metric of \textit{mean reciprocal rank} (MRR), which is the inverse of the position of the first relevant tuple in the list of matched entities. The local data source uses this feedback to modify its strategy and propagates them to the external data sources so they can adapt their strategies according to the feedback.

\section{Learning Mechanisms}
\label{sec:learning}

The main dilemma of online learning is to balance \textit{exploiting} the information known so far to deliver accurate results in the short run and \textit{exploring} new actions that have \textit{not} been tried before to gain more knowledge and eventually learn a more accurate model in the long run. If an online learning method focuses on the former, it might \textit{not} improve its model significantly over time. 
Popular algorithms to balance exploration and exploitation in information systems, such as \textit{UCB-1}, assume the environment is static, where the distribution of (positive) feedback for each action is fixed. This assumption does \textit{not} hold in our setting, as both local and external data sources may learn to revise their strategies of submitting queries and returning answers. Thus, the success of an external query may vary over time.

{\bf Learning: in Dynamic Settings:} We use an online learning algorithm called \textit{Roth and Erev}, which is effective in both static and dynamic environments~\cite{roth1995learning,mccamish2018data}. Our modified Roth and Erev algorithm implicitly reinforces probabilities proportional to user feedback and selects queries to send or results to return using their respective probabilities. 
Given $L(t)$ is the local data sources strategy at time $t$, we have
$L_{ij}(t+1) =$ $\frac{S_{ij}(t+1)}{\sum\limits_{j'}^n S_{ij'}(t+1)}$ where $S_{ij}(t)$ in matrix $S(t)$ maintains the accumulated reward of using keyword query $q_j$ to express the entity and query pair $e_i$ over the course of interaction up to round $t$. $S(t)$ is updated using the following formula in which $r$ is the value associated with user feedback (MRR) and $\alpha$ is the learning rate.
\vspace{-0.5em}
\begin{equation}
	S_{ij}(t+1) =
	\begin{cases}
	    S_{ij}(t) + \alpha r & \text{User selected}\\
	    S_{ij}(t)	& \text{Otherwise}
	\end{cases}
	\label{eq:reinforce}
\end{equation}
\vspace{-1em}


A similar learning method is used for external data sources.

\textbf{Autonomous Communication:}
Learning only via new user feedback is costly as it prolongs user supervision and users may have to wait for a relatively long time to receive sufficiently accurate results. To alleviate this problem, ProgMap stores the user feedback on each interaction. Then, it starts a series of communications between the local and external data sources for the same query with the goal of moving the relevant answer(s) in the returned list of answers to higher position(s). Since we already know the relevant answers in these series of communication, it does \textit{not} need any additional user supervision. Once some minimum number of interactions completes, these communications continue until there is no improvement in the position of the relevant answers in the result list. It is also possible to perform these interactions at any time after a user interaction.

\textbf{Query Expansion:}
To improve the understandability of the crafted keyword queries by the local data source, ProgMap expands the keywords available to the local data source by using the returned tuples from the external one. If user feedback indicates that a returned tuple matches the desired tuple, the local data source converts that returned tuple into its keyword parts and adds them to its strategy.

\begin{figure}[h!]
	\vspace{-1em}
	\centering
	\includegraphics[width = 0.9\linewidth]{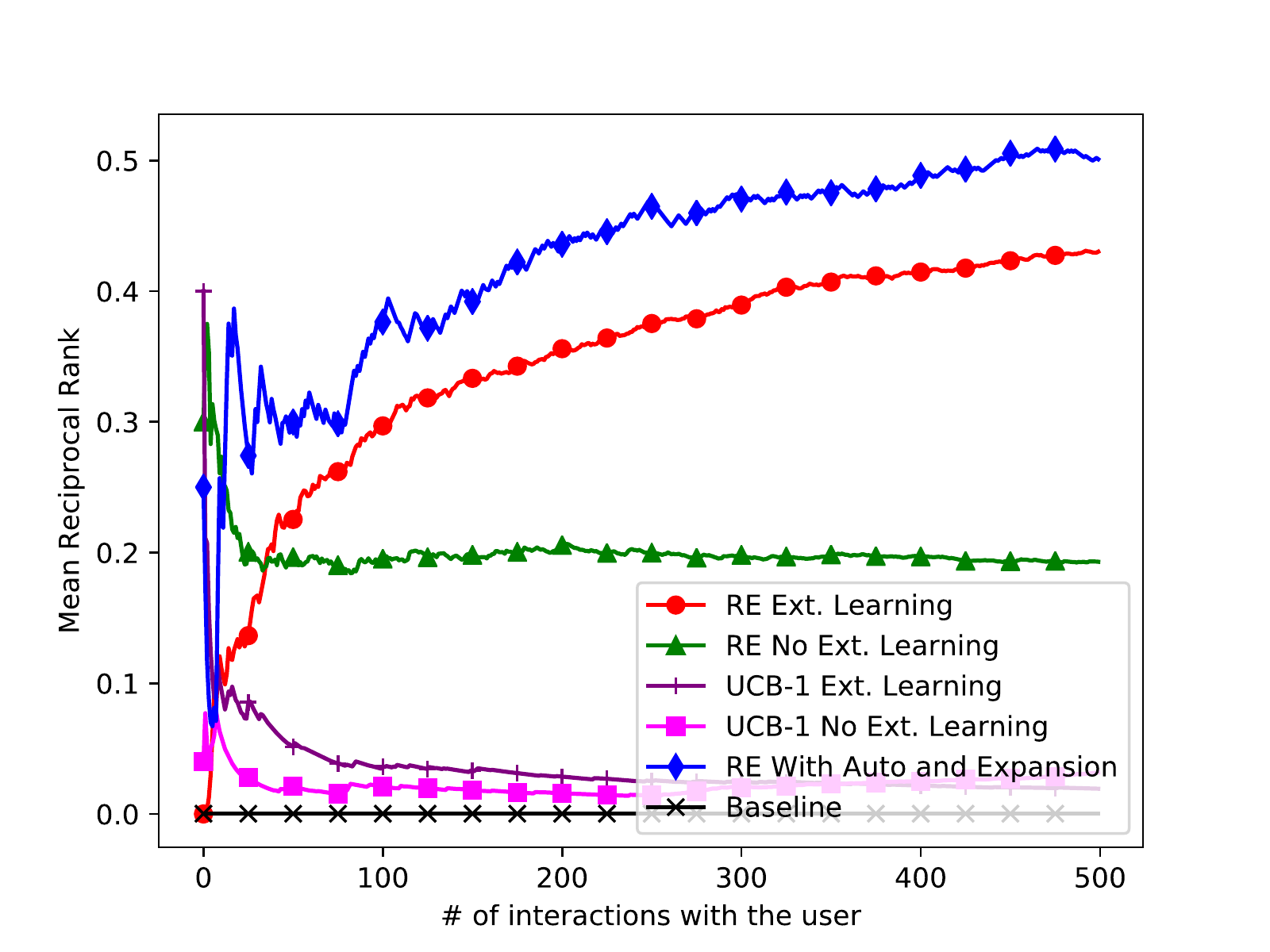}
	\vspace{-1em}
	\caption{MRR Over Product Datasets}
	\label{fig:results:ag}
	\vspace{-1em}
\end{figure}

\begin{figure}[h!]
	\vspace{-0.5em}
	\centering
	\includegraphics[width = 0.9\linewidth]{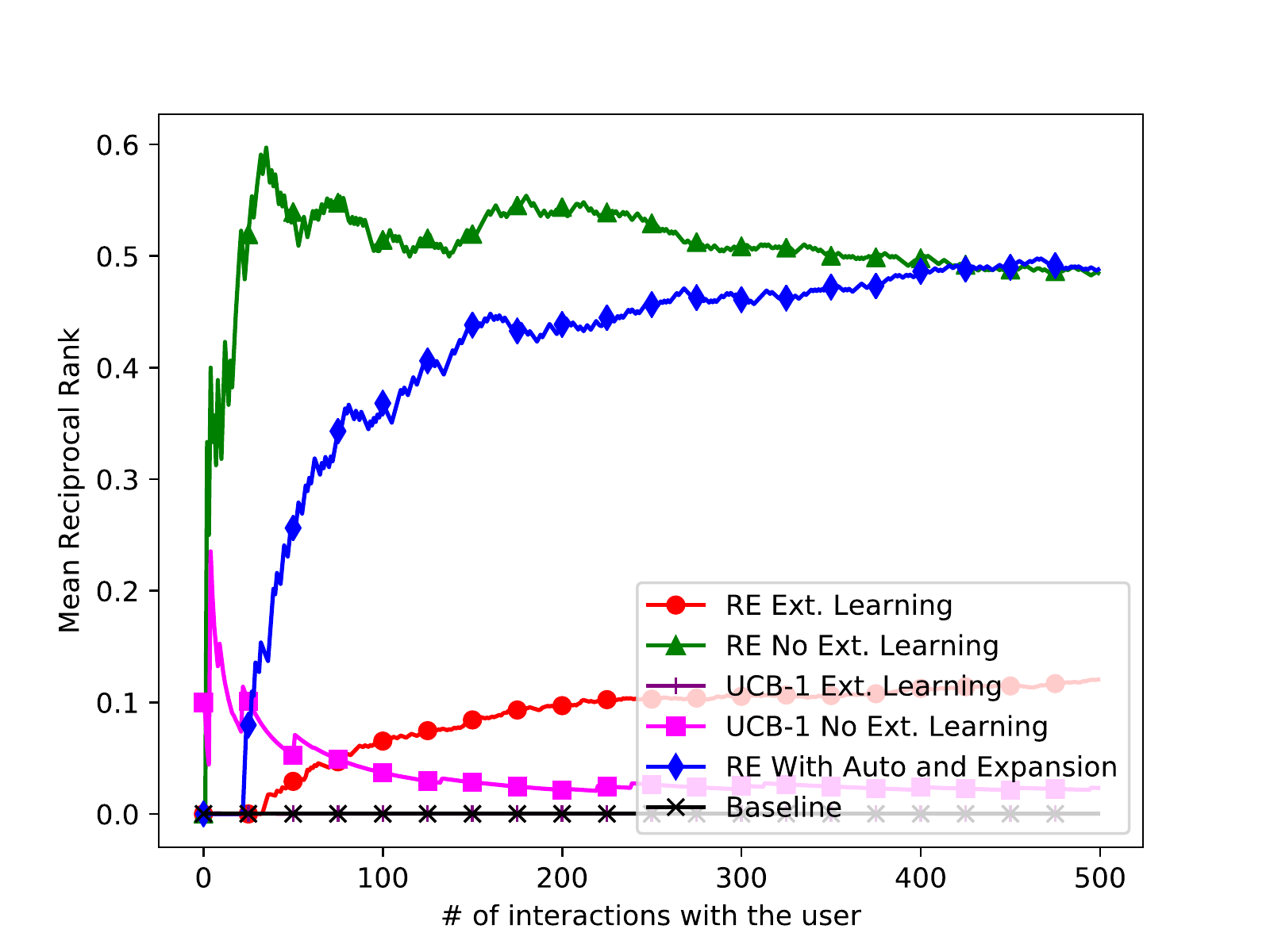}
	\vspace{-1em}
	\caption{MRR Over Movie Datasets}
	\label{fig:results:movie}
	\vspace{-1em}
\end{figure}

\textbf{Empirical Study:}
We evaluated our proposed method and several other methods of collaboration between local and external data sources using two real-world sets of databases: subsets of Amazon and Google product databases from \url{https://dbs.uni-leipzig.de/en/research/projects/object_\\matching}) consisting of approximately 5,000 tuples in each database, and
two different movie datasets from \url{https://sites.google.com/site/anhaidgroup/useful-stuff/data} each consisting of approximately 100,000
tuples. Figures~\ref{fig:results:ag} and~\ref{fig:results:movie} illustrate our results over the product and movie databases respectively. Each external data source returns 20 results.
Baseline is the result of the local data source sending every keyword for a given intent and the external returning entities according to BM25 ranking, a popular ranking method for keyword queries. We have evaluated the results of our modified Roth and Erev algorithm, ``RE'' in the figures, and UCB-1 for adapting the local strategy. We also evaluate the case where the external data source does \textit{not} learn and uses BM25 ranking, shown as ``No Ext. Learning'', and when it adapts it uses the same learning algorithm as the local one, shown as ``Ext Learning''. ``RE with Auto and Expansion'' is the method where both data sources learning according to our algorithm with the proposed optimization techniques. The results on both datasets indicate that our system outperforms other methods and delivers reasonably accurate results after only about 100 interactions with the user. The accuracy of our system increases over time, e.g., MRR of 0.75 after 2000 interactions (not shown in the figure). The method in which the external data source does \textit{not} learn may initially perform better than our system, e.g., Figure~\ref{fig:results:movie}, but after a while our system catches up with and outperforms it.


	\vspace{-0.5em}
	\vspace{-1em}

\bibliographystyle{abbrv}
\bibliography{ref.bib}

\end{document}